\documentstyle[12pt,aasms4]{article}

\slugcomment{To appear in Astrophysical Journal}

\lefthead{Weiler et al.}  
\righthead{RSNe as Distance Indicators}

\begin{document}

\title{Radio Supernovae as Distance Indicators}
\author{Kurt W. Weiler} 
\affil{Remote Sensing Division, Naval Research
Laboratory, Code 7214, Washington, DC 20375-5320;
http://rsd-www.nrl.navy.mil/7214/weiler/; kweiler@SNe.nrl.navy.mil}

\author{Schuyler D. Van Dyk\altaffilmark{2}} 
\affil{Dept. of Physics \& Astronomy, UCLA, Los Angeles, CA 90095-1562;
vandyk@astro.ucla.edu}

\author{Marcos J. Montes\altaffilmark{1}} 
\affil{Remote Sensing Division, Naval
Research Laboratory, Code 7214, Washington, DC 20375-5320;
montes@rsd.nrl.navy.mil}

\author{Nino Panagia\altaffilmark{3}} 
\affil{Space Telescope Science
Institute, 3700 San Martin Drive, Baltimore, MD 21218;
panagia@stsci.edu}

\and

\author{Richard A. Sramek} 
\affil{P.O. Box 0, National Radio Astronomy
Observatory, Socorro, NM 87801; dsramek@nrao.edu}

\altaffiltext{1}{Naval Research Lab/National Research Council
Cooperative Research Associate.}
\altaffiltext{2}{Visiting scientist.}  
\altaffiltext{3}{Affiliated with the Astrophysics Division, Space
Science Department of ESA.}

\begin{abstract}

Long-term monitoring of the radio emission from supernovae with the
Very Large Array (VLA) shows that the radio ``light curves'' evolve 
in a systematic
fashion with a distinct peak flux density (and thus, in combination
with a distance, a peak spectral luminosity) at each frequency and a
well-defined time from explosion to that peak.  Studying these two
quantities at 6 cm wavelength, peak spectral luminosity ($L_{\rm 6\ cm\
peak}$) and time after explosion date ($t_0$) to reach that peak
($t_{\rm 6\ cm\ peak} - t_0$), we find that they appear related.  In
particular, based on two objects, Type Ib supernovae may be
approximate radio ``standard candles'' with a 6 cm peak luminosity
$L_{\rm 6\ cm\ peak} \approx 19.9 \times 10^{26}$ erg s$^{-1}$ Hz$^{-1}$;
also based on two objects, Type Ic supernovae may be approximate radio
``standard candles'' with a 6 cm peak luminosity $L_{\rm 6\ cm\ peak}
\approx 6.5 \times 10^{26}$ erg s$^{-1}$ Hz$^{-1}$; and, based on
twelve objects, Type II supernovae appear to obey a relation $L_{\rm 6\ cm\
peak} \simeq 5.5 \times 10^{23} (t_{\rm 6\ cm\ peak} - t_0)^{1.4}$ erg
s$^{-1}$ Hz$^{-1}$, with time measured in days.  If these relations
are supported by further observations, they provide a means for
determining distances to supernovae, and thus to their parent
galaxies, from purely radio continuum observations.

With currently available sensitivity of the VLA, it is possible to
employ these relations for objects further than the Virgo Cluster out
to $\sim$ 100 Mpc.  With planned improvements to the VLA and the
possible construction of more sensitive radio telescopes, these
techniques could be extended to $z \sim 1$ for some classes of bright
radio supernovae.

\end{abstract}


\section{Introduction}

A series of papers published over the past 15 years on radio
supernovae (RSNe; see references to Table 2) have established the
radio evolution for 16 objects: 2 Type Ib supernovae (SN 1983N, SN
1984L), 2 Type Ic supernovae (SN 1990B, SN 1994I) and 12 Type II
supernovae (SN 1970G, SN 1978K, SN 1979C, SN 1980K, SN 1981K, SN
1985L, SN 1986E, SN 1986J, SN 1987A, SN 1988Z, SN 1993J, and SN
1996cb).  In this extensive study of the radio emission from
supernovae (SNe), in most cases using the Very Large Array 
(VLA)\footnote{The VLA is a telescope of the National Radio Astronomy
Observatory which is operated by Associated Universities, Inc., under
a cooperative agreement with the National Science Foundation.} radio
telescope, two effects have been noticed: 1) Type Ib and Type
Ic SNe have roughly constant 6 cm radio luminosities at peak with only
slight differences between the two types, and 2) Type II SNe appear to
have a higher 6 cm radio luminosity at peak if they take longer to
reach that peak.  These two effects are also consistent with the
tenets of the SN shock/circumstellar medium interaction model of
Chevalier (1982a, b; 1984) for the origin of the radio emission if: 1)
the radio emission for the rather homogeneous classes of Type Ib/c SNe
arises in similar circumstellar environments, and 2) the radio
emission for the more inhomogeneous class of Type II SNe arises in
diverse circumstellar environments, but the denser the circumstellar
medium the longer it takes for the material to become optically thin
to the radio emission and the brighter the radio emission.  If these
effects can be quantified, they can provide a purely radio-based
secondary distance indicator for SNe and, by association, for their
parent galaxies.  The RSNe available for study and some relevant
properties of their parent galaxies are listed in Table 1.

\section{Models}

While the relations to be described are empirical, there is some support
from theoretical modelling.  All known RSNe appear to share common
properties of: (a) nonthermal synchrotron emission with high
brightness temperature; (b) a decrease in absorption with time,
resulting in a smooth, rapid turn-on first at shorter wavelengths and
later at longer wavelengths; (c) a power-law decline of the emission
flux density with time at each wavelength after maximum flux density
(absorption $\tau \approx 1$) is reached at that wavelength; and, (d)
a final, asymptotic approach of spectral index $\alpha$ to an
optically thin, nonthermal, constant negative value (Weiler et
al. 1986, 1990). Chevalier (1982a, b) has proposed that the
relativistic electrons and enhanced magnetic field necessary for
synchrotron radio emission are generated by the outgoing shock wave
from the SN explosion interacting with a relatively high-density
envelope of ionized circumstellar material surrounding the
presupernova star.  This dense cocoon is presumed to arise from mass
loss in a stellar wind from a red supergiant SN precursor, or a
companion, which has been ionized and heated by the initial UV/X-ray
flash of the SN explosion.  This circumstellar material (CSM) is also
the source of the initial absorption.  A rapid rise in the observed
radio flux density results from the shock overtaking more and more of
the wind material, leaving progressively less of it along the line of
sight to the observer to absorb the slowly decreasing synchrotron
emission from the shock region.

The radio light curve which results from these two competing effects
of rapidly declining absorption and more slowly declining emission is
shown schematically for one frequency in Figure 1.

\subsection{Parameterized Radio Light Curves}

It has been shown by Weiler et al. (1986) that the radio emission from
RSNe can be described, for simple cases, by:

\begin{equation}
S {\rm (mJy)} = K_1 {\left({\nu} \over {\rm 5~GHz}\right)^{\alpha}}
{\left({t - t_0} \over {\rm 1~day}\right)^{\beta}} e^{-{\tau}},
\end{equation}

\noindent where

\begin{equation}
\tau = K_2 {\left({\nu} \over {\rm 5~GHz}\right)^{-2.1}} {\left({t -
t_0} \over {\rm 1~day}\right)^{\delta}},
\end{equation}

\noindent with $K_1$ and $K_2$ corresponding, formally, to the flux
density and uniform external absorption, respectively, at 5 GHz (6 cm
wavelength) one day after the date of explosion, $t_0$.  The term
$e^{-\tau}$ represents the attenuation of a medium that completely and
uniformly covers the emitting source (i.e., a ``uniform'' external
absorption) and the absorption is assumed due to purely thermal,
ionized hydrogen with frequency dependence $\nu^{-2.1}$.  The
parameter $\delta$ describes the time dependence of the optical depth
for the uniform external absorbing medium.  The emission from the RSN
is assumed to be nonthermal synchrotron radiation with spectral index
$\alpha$ and to be decreasing with time with index $\beta$.

\subsection{Model Interpretation and Predictions}

The parameters of Equations 1 and 2 can be interpreted in terms of the
Chevalier (1982a, b; 1984) model if the dense, external cocoon of
material is established by a constant mass-loss rate ($\dot M$),
constant velocity ($w$) wind from a red supergiant progenitor.  The
time dependence $\beta$ of the flux density decline is derived by
assuming equipartition between the magnetic and relativistic particle
energies, a constant ratio of those energies to the thermal post-shock
particle energy, a power-law density distribution for the unshocked
circumstellar material (CSM) of $\rho_{\rm CSM} \propto r^{-s}$, and a
density distribution for the unshocked ejecta of $\rho_{\rm ej}
\propto r^{-n}$.  The time dependence $\delta$ of the external uniform
optical depth is clearly a function of both the shock expansion index
$m$ ($R_{\rm shock} \propto t^m$) and the radial distribution of
the CSM [$\delta = -3m = -3{(n-3) \over (n-s)}$].

The Chevalier model then interprets the parameters of Eq.~1 
at any given frequency $\nu$ as:

\begin{equation}
K_1 \propto (\dot M/w)^{(\gamma-7+12m)/4} \ \ {\rm or} \ \ K_1 \propto
(\dot M/w)^{a},
\end{equation}

\noindent where

\begin{equation}
a = (\gamma - 7 +12 m)/4,
\end{equation}

\noindent and

\begin{equation}
\beta = -(\gamma +5-6m)/2;
\end{equation}

and of Eq.~2 as

\begin{equation}
K_2 \propto (\dot M/w)^{(5-3m)} \ \ {\rm or} \ \ K_2 \propto (\dot
M/w)^{c},
\end{equation}

\noindent where

\begin{equation}
c = 5-3m,
\end{equation}

\noindent and

\begin{equation}
\delta = -3m.
\end{equation}

\noindent The term $\gamma$ is the index of the energy spectrum of the
relativistic synchrotron electrons (which is related to the radio
spectral index $\alpha$ by $\gamma= -2\alpha+1$), and $m$ is the
shock expansion index described above.

Substituting these relations into Eq.~1 and including the distance
to the SN to convert observed flux density ($S_{\nu}$) into spectral
luminosity ($L_{\nu}$), we can write

\begin{equation}
L_{\nu} \propto (\dot M/w)^a (t - t_0)^{\beta} \exp[(\dot M/w)^c
(t-t_0)^{\delta}].
\end{equation}

\noindent This is a function which has a single peak when the optical
depth approaches unity, so that we can obtain a value for the
mass-loss rate/presupernova wind velocity ratio ($\dot M/w$) from
setting the first derivative equal to zero (i.e., $dL_{\nu}/d(t-t_0)
\equiv 0$) which gives

\begin{equation}
(\dot M/w)^c \propto \tau_{peak} = \beta / \delta
\end{equation}

\noindent or

\begin{equation}
\dot M/w \propto (\beta / \delta)^{1/c} (t_{\nu\ {\rm peak}} -
t_0)^{-\delta/c}.
\end{equation}

\noindent Substituting this back into Eq.~9 yields a luminosity for
frequency $\nu$ at time $t_{\nu\ {\rm peak}}$ of:

\begin{equation}
L_{\nu\ {\rm peak}} \propto (\dot M / w)^a (t_{\nu\ {\rm peak}} -
t_0)^{\beta} {\rm e^{-\tau_{\nu\ {\rm peak}}}},
\end{equation}

\noindent or

\begin{equation}
L_{\nu\ {\rm peak}} \propto (\beta / \delta)^{a/c} {\rm e^{-(\beta /
\delta)}} (t_{\nu\ {\rm peak}} - t_0)^{\eta}
\end{equation}

\noindent where

\begin{equation}
\eta = -(a\delta - {\beta}c)/c.
\end{equation}

Eq.~13 implies that the peak luminosity at any frequency is related
to the time interval after explosion required for the radio emission
to reach that peak.  This is illustrated schematically in Figure 2.

In principle, the index $\eta$ in this simplified modelling treatment
is fully determined by only two physical parameters, the index of the
electron energy spectrum $\gamma$ (where $\gamma = -2\alpha + 1$) and
the index of the shock expansion $m$ (i.e., $R_{\rm shock} \propto
t^m$).  In practice, we do not expect to be able to describe the
evolution of a SN explosion so simply and will employ an empirical fit
to the available data. However, it is interesting to note that,
although there is considerable scatter, the simple model prediction
for the four Type Ib/c SNe gives, on average, relatively little
variation of spectral luminosity with turn-on delay [$L_{\nu\ {\rm
peak}} \propto (t_{\nu\ {\rm peak}}-t_0)^{\eta}$, $\eta \simeq -0.2$],
and for the twelve Type II SNe, a much stronger dependence of spectral
luminosity with turn-on delay ($\eta \simeq +1.4$).  For the Type II
SNe, it is quite surprising that this is the same index value as is
obtained from the actual fit to the data in $\S$4.2, Eq.~17.  However,
since the scatter in $\eta$ values is quite large, this
\underline{exact} agreement is certainly fortuitous even though the
indication of a steep slope is supported.

\subsection{Type II Peculiar SNe}

It should be noted that fitting the so-called Type II ``peculiar''
(Type IIpec), or Type IIn, RSNe 1986J (Weiler, Panagia, \& Sramek 1990)
and 1988Z (Van Dyk et al. 1993b), and Type IIb RSN 1993J (Van Dyk et
al. 1994) radio light curves requires additional absorption terms
beyond those given by Eqs.~1 and 2.  In particular, an additional term
describing an ``internal,'' or ``clumpy external,'' component of the
absorption, which is not included in the standard Chevalier model,
must be used and could imply that some Type II RSNe must be considered
separately.  However, since our results are empirically determined,
and since RSNe 1986J, 1988Z, and 1993J appear to obey the same
relation as their more normal brethren, they are included along with
other Type II SNe for the present discussion.
 
\subsection{SN 1987A}

One of the major peculiarities of SN 1987A is that its explosion
occurred when its progenitor was a blue supergiant (BSG) than being a
red supergiant (RSG).  Because of the much higher wind velocity of a
BSG than that of a RSG, the density of the CSM, i.e., the density of
the presupernova stellar wind ($\propto \dot M/w$), around SN 1987A was
considerably lower than around ``normal'' Type II SNe and, therefore,
its early radio emission was intrinsically much weaker and evolved
much more quickly than normal.  In particular, the flux density at 6
cm reached a peak only 1 day after the explosion and fell below
detection limits only a few days later.  As a consequence, despite the
efforts of Australian radio astronomers, the radio light curve of SN
1987A is poorly defined.

This fact, and the large difference in circumstellar environment (much
lower density, much higher velocity) as compared with normal Type II
SNe, makes it uncertain whether a quantitative comparison of SN 1987A
with the other Type II SNe can be made.  However, it is apparent in
Fig.~3 that SN 1987A agrees reasonably well with an extrapolation of
the relation determined for the remaining Type II SNe which have much
greater peak 6 cm radio luminosities and much longer times from
explosion to 6 cm peak flux density.  This may imply that the relation
is indeed valid for all Type II RSNe and that the gap between RSN 1996cb and
RSN 1987A is due to selection effects against fast turn-on, low radio
luminosity RSNe.  Thus, since we are dealing with an empirical
relation, we have included SN 1987A in our discussion.

\section{Data}

\subsection{Peak Luminosity and Time to Peak}

In order to quantify possible relations between SN type, peak radio
luminosity, and time from explosion to radio peak, it is necessary to
establish these quantities in a systematic fashion. The data sets
available for the various RSNe are of quite variable quality, and most
RSNe show some deviation from smooth radio light curves.  Thus, even
when observations are available at 6 cm wavelength near the time of
peak flux density, the smoother best-fit model values have been used 
in all cases
to determine the two quantities of interest: peak 6 cm flux density,
$S_{\rm 6\ cm\ peak}$ (from which the peak 6 cm spectral luminosity,
$L_{\rm 6\ cm\ peak}$, is calculated using the distances given in Table
1), and the time in days ($t_{\rm 6\ cm\ peak} - t_0$) from explosion,
$t_0$, to 6 cm peak luminosity, $t_{\rm 6\ cm\ peak}$.  These values,
along with associated errors, are given in Table 2 for each of the 16
RSNe under consideration.

Two caveats should be kept in mind.  First, new fits have been
performed for all available data sets and, due to revisions or the
inclusion of new data, some fitting parameters may differ slightly
from previously published values.  Second, these results are from
preliminary fits to some of the data sets and more data may somewhat
alter the parameter values in future treatments.

\subsection{Errors}

Because of the very diverse nature of the data, missing data at
critical (usually early) times, and variations in the assumptions for
the modelling of different RSNe, it is extremely difficult to assign
errors to the basic quantities of S$_{\rm 6\ cm\ peak}$, including, in
many cases, large uncertainty in determining or assigning a date of
explosion, t$_0$.  Additionally, the conversion of S$_{\rm 6\ cm\ peak}$
into L$_{\rm 6\ cm\ peak}$ introduces the additional uncertainty of poor
distance estimates to many of the parent galaxies of the SNe.

However, to try to provide some indication of the relative quality of
the different points, we have attempted to assign reasonable error
estimates to all values.  For the distances listed in Table 1, we have
tried to find the best value and error estimate available in the
literature and, when successful, have listed the appropriate
reference.  For galaxies where no estimate is available, we have taken
the value from Tully (1988), accepting all his corrections but
adjusting his distance value obtained with his assumed H$_0$ = 75 km
s$^{-1}$ Mpc$^{-1}$ to what appears to be a more recently preferred
value of H$_0$ = 65 km s$^{-1}$ Mpc$^{-1}$ (see, e.g., Saha, et
al. 1997 and references therein; Riess, Press, \& Kirshner 1996;
Hamuy, et al. 1996).  Since Tully (1988) assigns no distance error to
his values, we have assumed that $\pm$ 15 \%, to roughly cover the
range of H$_0$ from 55 to 75 km s$^{-1}$ Mpc$^{-1}$ within one sigma,
is a reasonable estimate.

For the other two critical parameters, S$_{\rm 6\ cm\ peak}$ and 
(t$_{\rm 6\ cm\ peak}$ - t$_0$), we have used several methods to obtain rough 
error estimates.

Where enough data are available and the radio light curves are well
determined (e.g., SN 1993J with hundreds of measurements), we were
able to use the bootstrap method\footnote{Bootstrap (Press et
al.~1992) procedures use the actual data sets to generate thousands of
synthetic data sets that have the same number of data points, but some
fraction of the data is replaced by duplicated original points. The
fitting parameters are then estimated for these synthetic data sets
using the same algorithms that are used to determine the parameters
from the actual data. The ensemble of parameter fits is then used to
estimate errors for the parameters by examining the number distribution of
the parameter in question. The errors we present correspond to the
15.85\% and 84.15\% points in the distribution of the occurrence of
parameters from the synthetic data.}  to obtain direct estimates of
the one sigma deviation for both quantities.  However, in the cases
where the data were very sparse (e.g., SN 1986E with only one
detection at one frequency), we had to rely on a qualitative estimate
of the error by adjusting fitting parameters by hand, to obtain a
range of values for $S_{\rm 6\ cm\ peak}$ and ($t_{\rm 6\ cm\ peak} -
t_0$) which appeared consistent with the data.  In extreme cases where
one or both parameters were effectively indeterminate, that has been
indicated in Table 2 by the description ``ind.'' and, in Figure 3, by
a line stub for that side of the range of uncertainty.

Once error estimates were established for S$_{\rm 6\ cm\ peak}$ and the
distance, determination of error estimates for the L$_{6\rm \ cm\ peak}$
followed standard error propagation procedures.

Finally, error estimates for the ``radio distances'' in Table 3 were
obtained for Type Ib and Type Ic SNe from propagation of the flux
density error from Table 2 and the deviation from the average
luminosity (Equation 15 for Type Ib and Equation 16 for Type Ic) for
each SN type.  For Type II SNe errors were estimated from the flux
density error from Table 2 and the standard deviation of the
luminosities from the straight line shown in Figure 3.  While it is
felt that these error estimates are reasonable for the quality of the
data available, they cannot be rigorously justified.

\section{Discussion}

\subsection{Type Ib/c Peak Radio Luminosities and Distances}

The model discussion presented in Section 2 suggests that Type Ib/c
SNe may have relatively little variation in their peak spectral
luminosities, even if there are variations in the measured time from
explosion to peak.  Unfortunately, there are only four Type Ib/c RSNe
which have been studied at radio wavelengths, and even these four
examples are diverse, with two being Type Ib and two being Type Ic.
Additionally, SN 1984L has very limited data available, and SN 1990B
was quite distant and therefore studied with poor signal-to-noise.
Within these limitations, however, examination of Table 2 
shows that Type Ib RSNe, based on data for only two
objects, may be approximate standard candles with peak 6 cm
luminosities ranging only from 14.1 -- 25.7 $\times$ 10$^{26}$ erg
s$^{-1}$ Hz$^{-1}$, and, similarly, Type Ic RSNe, again based on data
for only two objects, may be approximate standard candles with peak 6
cm luminosities ranging only from 5.6 -- 13.7 $\times$ 10$^{26}$ erg s$^
{-1}$ Hz$^{-1}$.  These are consistent to within the errors for Type
Ib SNe having an average peak 6 cm luminosity of

\begin{equation}
L_{\rm 6\ cm\ peak} \approx 19.9 \times 10^{26} \ {\rm erg}\ {\rm
s}^{-1}\ {\rm Hz}^{-1}
\end{equation}

\noindent and Type Ic SNe having an average peak 6 cm luminosity of

\begin{equation}
L_{\rm 6\ cm\ peak} \approx 6.5 \times 10^{26} \ {\rm erg}\ {\rm
s}^{-1}\ {\rm Hz}^{-1}.
\end{equation}

If we assume that Type Ib and Type Ic RSNe are, in fact, standard
candles in their peak 6 cm radio emission and that, on average, the
assumed distances listed for the objects in Table 1 are correct, we
can determine a ``radio distance'' to each RSN, such that its measured
peak 6 cm flux density would yield the average peak 6 cm spectral
luminosity for its particular type.  These distances are listed in
Table 3 along with an error.  The ``assumed distances'' and
errors from Table 1 are repeated in Table 3 for the
reader's convenience.

While it should be kept in mind that the two sets of distance
estimates are not completely independent, with the assumed
``independent'' distances {\it de facto} setting the scale for the
radio distances, it is apparent that the data are consistent, to
within the errors, with an assumption of constant 6 cm peak
luminosity.  Thus, the distance to any Type Ib or Type Ic RSN can be
estimated simply by measuring its 6 cm peak
flux density and comparing this observed value with
the luminosity predicted by Eqs.~15 and 16.
Clearly, more objects of both subtypes must be identified and studied
in the radio to better test and define these two equations.

\subsection{Type II Peak Radio Luminosities and Distances}

There are 12 examples of Type II RSNe which have a sufficient number
of radio observations available for detailed study.  While Type II
RSNe are far more heterogeneous in their radio (as well as in their
optical) properties than Type Ib/c RSNe, inspection of Table 2 and
Figure 3 shows that the longer it takes a Type II RSN to reach 6 cm
peak flux density, the higher its 6 cm luminosity at that peak.  This
is also supported by the model discussion in Section 2, which implies
a relatively steep power-law relation between the two quantities.
However, as with the Type Ib/c SNe, we are considering here an
empirical test and therefore will determine a best-fit power-law to
the available turn-on time {\it vs.\/} peak 6 cm luminosity data.

Performing an unweighted least squares fit to the data for the 12
available Type II RSNe, we obtain the relation (formal fitting errors
given) of

\begin{equation}
L_{\rm 6\ cm\ peak} \simeq 5.5 {+8.7 \choose -3.4} \times 10^{23} (t_{\rm 6\
cm\ peak} - t_0)^{1.4 \pm 0.2} \ {\rm erg}\ {\rm s}^{-1}\ {\rm
Hz}^{-1},
\end{equation}

\noindent with time in days.  This relation is shown as the {\it
dashed\/} line in Figure 3.  We note that the empirically derived
slope of the $L_{\rm 6\ cm\ peak}$ {\it vs.\/} ($t_{\rm 6\ cm\ peak} -
t_0$) relationship agrees well with the value inferred from
model estimates (see Eq. 14).  This result suggests the validity of
the Chevalier (1982a,b) model, at least for interpreting
the origin of the radio emission.

If we assume that Type II RSNe obey this relation and that, on
average, the independent distances listed for the objects in Table 1
are correct, we can determine a ``radio'' distance to each RSN, such
that its measured peak 6 cm flux density would yield the ``best-fit''
6 cm spectral luminosity predicted by Eq.~17.  These ``radio distances''
are listed in Table 3, Column 4, along with errors, while the
``assumed distances'' and errors from Table 1 are repeated in
Table 3 for the reader's convenience.

Again, while it should be kept in mind that the two sets of distance
estimates are not completely independent, with the ``assumed distances''
 {\it de facto} setting the scale for the
radio distances, it is apparent that the data are consistent, to
within the errors, with the results of Eq. 17.  Thus, the distance
to any Type II RSN can be estimated simply by measuring its time from
explosion to 6 cm peak flux density and comparing the observed 6 cm
peak flux density with the predicted luminosity.

\section{Deviations}

Examination of the distances determined from the relations of
Eqs.~15, 16, and 17 for Type Ib, Type Ic and Type II,
respectively, in Table 3 shows that they are generally in
good agreement to within the estimated errors.  However, several cases show 
relatively large disagreement inconsistent with the error estimates.

\subsection{SN 1985L, SN 1986E, and SN 1996cb}

Perhaps too easily, large deviations for these three SNe can be
explained, if not dismissed.  SN 1985L was faint (S$_{\rm 6\ cm\ peak}$
$\sim$ 0.65 mJy, estimated) and had only two clear detections of
radio emission (Van Dyk et al., 1998).  While reasonable estimates of
RSN parameters can be obtained for SN 1985L through the use of a number of upper
limits, the quality of the $L_{\rm 6\ cm\ peak}$ and ($t_{\rm 6\ cm\ peak} -
t_0$) must remain suspect.

SN 1986E (Montes et al.~1997) has the same difficulties as SN 1985L, 
but even more extreme.
Only one clear radio detection at one frequency exists, and the rest of
the parameter fitting has to rely on the use of upper limits.  It was
also quite faint, with a peak 6 cm flux density of only 
S$_{\rm 6\ cm\ peak} \sim$ 0.31 mJy.

SN 1996cb (Van Dyk et al., in preparation), while relatively bright 
with a 6 cm peak flux density of S$_{\rm 6\ cm\ peak} \sim$ 1.8 mJy,
exhibited the most rapid turn-on of any Type II RSNe, except for SN 1987A.
(In fact, many of its radio properties resemble those of the Type Ib/c
RSNe.)
This, unfortunately, meant that the turn-on at wavelengths shorter 
than 20 cm was completely missed, and estimates of $S_{\rm 6\ cm\ peak}$
and ($t_{\rm 6\ cm\ peak} - t_0$) are rather uncertain.

\subsection{SN 1980K in NGC 6946}

RSN 1980K (Weiler et al.~1986, 1992)
has a very large deviation from Eq.~17 (the {\it dashed\/} line in
Figure 3).  Also, it is possibly the most
interesting deviation in that it is not easily explained away.  SN
1980K was a reasonably strong, well-studied, ``normal'' Type II RSN,
where there can be little doubt as to its characteristics and model
parameters.  Thus, it may well serve as a test case.

If the relation of Eq.~17 for Type II SNe is valid, and SN 1980K is
indeed a normal Type II RSN, then the distance estimate of 6.3 Mpc
from Tully (1988; converted to $H_0$ = 65 km s$^{-1}$ Mpc$^{-1}$) for
the parent galaxy, NGC 6946, must be too short.  A significantly
larger distance of $\gtrsim$ 10 Mpc is implied by the radio results.

NGC 6946 is a large ($11.5 \times 9.8$ arcmin), nearby ($v_{\rm
heliocentric} = 46$ km\ s$^{-1}$; Tully 1988) spiral (Scd) galaxy,
which has been a prolific producer of SNe in the last century (SNe
1917A, 1939C, 1948B, 1968D, 1969P, and 1980K).  However, it is at a
low galactic latitude ($b = 11.7^{\circ}$) and located along the
direction of rotation of the Milky Way ($l = 95.7^{\circ}$), requiring
large corrections to obtain an intrinsic radial velocity with respect
to the galactic rest frame.  Because it is so close, peculiar motions
within the Local Group of galaxies may also disturb the Hubble flow
and yield an inaccurate distance estimate from corrected radial
velocity measurements.

Recent work on the Expanding Photosphere Method (EPM) by Schmidt,
Kirshner, \& Eastman (1992) yields a distance to SN 1980K (and thus,
NGC 6946) of $8.1 \pm 1.5$ Mpc from IR observations and 7.2 (+0.7, 
$-1.0$) Mpc from optical observations, both somewhat closer than our
radio estimate of $12.3 \pm 1.9$ Mpc.  Furthermore, a revision of the EPM
results by Schmidt et al.~(1994) yields a distance of $5.7 \pm 0.7$
Mpc, in much closer agreement with the Tully (1988) distance than with
our radio distance.

We have circumstantial evidence that NGC 6946 may be more distant than
commonly assumed.  The fact that NGC 6946 is the galaxy with the
largest number of SN events ever recorded suggests a high blue
luminosity, which should be at least as high as that of another
prolific SN producer, NGC 4321 (M100), in which 4 SNe have been
discovered (with only one, SN 1979C, being a {\it bona fide} Type II
SN).  If we assume that NGC 6946 has the same blue luminosity, $M_B$,
as NGC 4321 and establish the difference in their total apparent 
magnitudes as $m_B$(NGC 4321) $-$ $m_B$(NGC 6946) = 0.47 (Tully 1988),
accepting that the difference of their reddening corrections is $\lesssim$1
magnitude (as indicated by optical and UV observations of SNe 1979C and 
1980K; Panagia 1982), we find that the distance to NGC 6946 is no less than
0.51 times the distance to NGC 4321 ($d = 17.1 \pm 1.3$ Mpc;
Freedman et al. 1994a).  Thus, this implies a distance to NGC 6946 of
$d > 8.7$ Mpc.

In any case, it appears that the distance to NGC 6946 is sufficiently
uncertain that it may be in error by the factor of $\sim$ 2 which the
radio data for SN 1980K imply.  Clearly, a measurement of the distance
to NGC 6946 with more accurate techniques (e.g., Cepheids) is called for 
and could provide an important test of the $L_{\rm 6\ cm\ peak}$ {\it vs.\/} 
$(t_{\rm 6\ cm\ peak} - t_0)$ hypothesis.

\section{Conclusions}

We have presented evidence that the radio emission from SNe may have
quantifiable properties which allow for distance determinations.  Type
Ib RSNe, based on a statistically very small sample of only two
objects, may be approximate radio ``standard candles,'' with with a 6
cm peak luminosity of $\sim 19.9 \times 10^{26}$ erg s$^{-1}$
Hz$^{-1}$; Type Ic RSNe, also based on a very small sample of only two
objects, may be approximate radio ``standard candles'' with a 6 cm
peak luminosity of $\sim 6.5 \times 10^{26}$ erg s$^{-1}$ Hz$^{-1}$;
and Type II RSNe, based on a sample of twelve objects, appear to obey
a relation $L_{\rm 6\ cm\ peak} \simeq 5.5 \times 10^{23}\ (t_{\rm 6\
cm\ peak} - t_0)^{1.4}$ erg s$^{-1}$ Hz$^{-1}$ (with time in days).
Thus, measurement of the radio turn-on time ($t_{\rm 6\ cm\ peak} -
t_0$) and peak flux density $S_{\rm 6\ cm\ peak}$ can yield a
luminosity estimate and therefore a distance.

The reality of the ``standard radio candle'' hypothesis for Type Ib
and Type Ic RSNe may be tested simply through the study of more
objects.  For Type II RSNe, the large deviation of the well-observed
SN 1980K in NGC 6946 from the distance predictions of this hypothesis
could serve as a useful test.  Whereas the presently accepted distance
is $\sim$6 Mpc to NGC 6946 (Tully 1988), the radio results predict a
significantly larger distance of $\gtrsim$10 Mpc.  A new, accurate
determination of the distance to NGC 6946 is certainly called for and,
in general, improved distances to all host galaxies of RSNe would be
valuable.

Although there are still relatively few objects to which these
techniques can be applied, RSNe may eventually provide a powerful and
independent technique for investigating the long-standing problem of
distance estimates in astronomy.  With such intrinsically bright Type
II RSNe as SN 1988Z and SN 1986J, the technique can be applied to
distances of at least 100 Mpc with current VLA
technology.  With future sensitivity improvements and planned, new,
more sensitive radio telescopes, the technique can be extended to
large distances even for less luminous RSNe.  With a sensitivity of 1
$\mu$Jy, a SN of the same class as SNe 1988Z and 1986J could be
detected up to a redshift of $\sim$ 1, while a more normal Type II SN,
such as SN 1980K, could be studied accurately up to a redshift of
$\gtrsim 0.1$.

\acknowledgments

KWW \& MJM wish to thank the Office of Naval Research (ONR) for the
6.1 funding supporting this research.  Additional information and data
on radio supernovae can be found on {\bf
http://rsd-www.nrl.navy.mil/7214/weiler/} and linked pages.

\clearpage

\begin{deluxetable}{llllrcc}
\tablewidth{450pt}
\tablecolumns{7}
\tablecaption{Radio Supernova List and Parent Galaxy Properties
\label{tbl-1}}
\tablehead{\colhead{RSN} & \colhead{Optical} & \colhead{Parent} &
\colhead{Hubble} & \colhead{Systemic} & \colhead{Assumed} &
\colhead{Dist.} \nl
\colhead{Name} & \colhead{Type} & \colhead{Galaxy} &
\colhead{Type} & \colhead{Velocity\tablenotemark{a}} & 
\colhead{Distance\tablenotemark{b}} & \colhead{Ref.\tablenotemark{c}} \nl
\colhead{} & \colhead{} & \colhead{} & \colhead{} & 
\colhead{(km\ s$^{-1}$)} & \colhead{(Mpc)} & \colhead{}}
\startdata
\sidehead{Type Ib/c}
1983N & Ib & NGC 5236 (M83)  & ScI--II & 337 & $5.4 \pm 0.8$ & \nl
1984L & Ib & NGC 991 & ScII  & 1541 & $21.7 \pm 3.3$ & \nl
1990B & Ic & NGC 4568 & Sbc  & 2168 & $19.4 \pm 2.9$ & \nl
1994I & Ic & NGC 5194 (M51)  & Sbc & 573 & $8.9 \pm 1.3$ & \nl
\sidehead{Type II}
1970G & IIL & NGC 5457 (M101) & Scd & 379 & $7.4 \pm 0.6$ & 1 \nl 
1978K & II & NGC 1313        & Sd & 254 & $4.5 \pm 0.7$ &  \nl
1979C & IIL & NGC 4321 (M100) & Sbc & 1522 & $17.1 \pm 1.8$ & 2 \nl
1980K & IIL & NGC 6946        & Scd & 338 & $6.3 \pm 1.0$ & \nl
1981K & II & NGC 4258 (M106)  & Sbc & 521 & $5.9 \pm 1.1$ & 3 \nl 
1985L & IIL & NGC 5033         & Sc & 931 & $21.6 \pm 3.2$ &  \nl
1986E & IIL & NGC 4302        & Sc & 1044 & $19.4 \pm 2.9$ &  \nl
1986J & IIn & NGC 891         & Sb & 712 & $11.1 \pm 1.7$ & \nl
1987A & IIpec & LMC       & Irr & 3.9 & $0.051 \pm 0.002$ & 4 \nl
\tablebreak
1988Z & IIn & MCG $+$03$-$28$-$022 & S & 6670 & $102.6 \pm 15.4$ & 5 \nl
1993J & IIb & NGC 3031 (M81)  & Sab & 96 & $3.6 \pm 0.3$ & 6 \nl
1996cb & IIb & NGC 3510       & SBm & 770 & $9.1 \pm 1.4$ & \nl
\enddata
\tablenotetext{a}{From Tully (1988); heliocentric velocity of the
parent galaxy is corrected for Galactic rotation of 300 km\ s$^{-1}$
towards $l=90^{\circ}$, $b=0^{\circ}$.}
\tablenotetext{b}{Unless otherwise referenced, the distance is from
Tully (1988), and is based on systemic velocity, corrected by a model
which assumes the Milky Way is retarded by 300 km\ s$^{-1}$ from
universal expansion by the mass of the Virgo Cluster, and on an
assumed Hubble constant of $H_0 = 65$ km\ s$^{-1}$ Mpc$^{-1}$ [Tully (1988) uses $H_0 = 75$ km\ s$^{-1}$ Mpc$^{-1}$].
Unless otherwise referenced, errors are conservatively
taken to be $\pm 15 \%$, a value which is large enough to encompass
both $H_0 = 55$ and $H_0 = 75$ km\ s$^{-1}$ Mpc$^{-1}$ within 1
$\sigma$.}
\tablenotetext{c}{(1) Kelson et al. ~1996; 
(2) Freedman et al.~(1994a); (3) average of VLBI
H$_2$O maser dynamical distances from Greenhill et al.~(1995) [$5.4
\pm 1.3$ Mpc] and from Miyoshi et al.~(1995) [$6.4 \pm 0.9$ Mpc]; 
(4) Panagia et al. 1991; (5)
systemic velocity from Stathakis \& Sadler (1991), and distance
derived from this velocity and an assumed $H_0 = 65$ km\ s$^{-1}$
Mpc$^{-1}$; (6) Freedman et al.~(1994b).}
\end{deluxetable}

\clearpage

\begin{deluxetable}{llcccc}
\tablewidth{500pt}
\tablecaption{Radio Supernova Properties
\label{tbl-2}}
\tablecolumns{6}
\tablehead{\colhead{RSN} & \colhead{Opt.} &
\colhead{Peak 6 cm} & \colhead{Time from Explosion} & \colhead{Peak 6
cm} & \colhead{RSN} \nl
\colhead{Name} & \colhead{Type} & \colhead{Flux Density\tablenotemark{a}} 
& \colhead{to 6 cm Peak\tablenotemark{a}} &
\colhead{Luminosity\tablenotemark{a,b}} & \colhead{Refs.\tablenotemark{c}} \nl 
\colhead{} & \colhead{} & \colhead{($S_{\rm 6\ cm\ peak}$)} & 
\colhead{($t_{\rm 6\ cm\ peak} - t_0$)} & \colhead{($L_{\rm 6\ cm\ peak}$)} 
& \colhead{} \nl
\colhead{} & \colhead{} & \colhead{(mJy)} & \colhead{(days)} & 
\colhead{($10^{26}$ erg s$^{-1}$ Hz$^{-1}$)} & \colhead{}}
\startdata
\sidehead{Type Ib/c}
1983N & Ib & 40.1 $+1.0 \choose -5.1$ & 11.6 $+1.0 \choose -1.0$ & 
14.1 $+4.2 \choose -4.6$ & 1,2 \nl
1984L & Ib & 4.6 $+{\rm ind.} \choose -4.0$ & 11.0 $+29.0 \choose -{\rm ind.}$ 
& 25.7 $+{\rm ind.} \choose -23.7$ & 1,3 \nl
1990B & Ic & 1.3 $+0.1 \choose -0.1$ & 37.5 $+1.7 \choose -1.6$ 
& 5.6 $+1.7 \choose -1.8$ & 4 \nl
1994I & Ic & 14.6 $+1.6 \choose -0.5$ & 35.8 $+2.0 \choose -2.2$ 
& 13.7 $+4.4 \choose -4.1$ & 5 \nl
\sidehead{Type II}
1970G & IIL & 21.5 $+1.4 \choose -5.1$ & 307.0 $+30.0 \choose -45.0$ 
& 14.0 $+2.5 \choose -4.0$ & 1,6,7,8 \nl
1978K & II & 229.0 $+51.0 \choose -44.0$ & 685.0 $+20.0 \choose -18.0$ 
& 55.3 $+20.7 \choose -19.7$ & 9,10 \nl
1979C & IIL & 7.3 $+0.2 \choose -0.2$ & 605.0 $+81.0 \choose -17.6$ 
& 25.3 $+5.4 \choose -5.4$ & 1,11,12 \nl
1980K & IIL & 2.5 $+0.1 \choose -0.1$ & 140.0 $+29.3 \choose -27.7$ 
& 1.2 $+0.4 \choose -0.4$ & 1,13 \nl
1981K & II & 5.1 $+{\rm ind.} \choose -2.1$ & 33.8 $+36.4~ \choose -{\rm ind.}$
& 2.1 $+{\rm ind.} \choose -1.2$ & 1,14,15 \nl
1985L & IIL & 0.7 $+0.1 \choose -0.2$ & 309.0 $+130.0 \choose -5.0$ 
& 3.6 $+1.2 \choose -1.4$ & 16 \nl
1986E & IIL & 0.3 $+0.1 \choose -0.1$ & 224.0 $+19.0 \choose -32.0$ 
& 1.4 $+0.4 \choose -0.5$ & 17 \nl
1986J & IIn & 136.0 $+4.4 \choose -4.8$ & 1150.0 $+85.0 \choose -21.0$ 
& 199.0 $+60.0 \choose -60.1$ & 18 \nl
1987A & IIpec & 91.4 $+113.5 \choose -27.5$ & 1.0 $+2.2 \choose -0.4$ 
& 0.003 $+0.004 \choose -0.001$ & 19 \nl
\tablebreak
1988Z & IIn & 1.9 $+0.1 \choose -0.1$ & 1420.0 $+79.0 \choose -112.0$ 
& 237.0 $+71.4 \choose -71.3$ & 20 \nl
1993J & IIb & 95.2 $+1.8 \choose -1.5$ & 180.0 $+4.1 \choose -5.0$ 
& 15.0 $+2.8 \choose -2.8$ & 21 \nl
1996cb & IIb & 1.85 $+0.2 \choose -0.1$ & 19.4 $+4.9 \choose -4.6$ 
& 1.8 $+0.6 \choose -0.6$ & 22,23 \nl
\enddata
\tablenotetext{a}{See $\S$3.2 for discussion of errors.}
\tablenotetext{b}{Distances are from Table 1 and peak flux densities are
from Column 3.}
\tablenotetext{c}{(1) Weiler et al.~(1986); (2) Sramek, Panagia, \&
Weiler (1984); (3) Panagia, Sramek, \& Weiler (1986); (4) Van Dyk et
al.~(1993a); (5) Rupen et al.~(1998); (6) Allen et al.~(1976); (7)
Marscher \& Brown (1978); (8) Cowan, Goss, \& Sramek (1991); (9) Ryder
et al. (1993); (10) Montes, Weiler, \& Panagia (1997); (11) Weiler et al.~(1981);
(12) Weiler et al.~(1991); (13) Weiler et al.~(1992); (14) van der
Hulst et al.~(1983); (15) Van Dyk et al.~(1992); (16) Van Dyk et al. 1998; (17) Montes et al. (1997); (18) Weiler,
Panagia, \& Sramek (1990); (19) Turtle et al. (1987); (20) Van Dyk et
al.~(1993b); (21) Van Dyk et al.~(1994); (22) Van Dyk et al. (1996);
(23) K.~W. Weiler, private communication.}
\end{deluxetable}

\clearpage

\begin{deluxetable}{llcc}
\tablewidth{450pt}
\tablecolumns{4}
\tablecaption{Distances \label{tbl-3}}
\tablehead{\colhead{RSN} & \colhead{Optical} &
\colhead{Assumed} & \colhead{Radio} \nl
\colhead{Name} & \colhead{Type}
& \colhead{Distance\tablenotemark{a}} &
\colhead{Distance\tablenotemark{b}} \nl 
\colhead{} & \colhead{} & \colhead{(Mpc)} & \colhead{(Mpc)}}
\startdata
\sidehead{Type Ib/c}
1983N & Ib & $~\,5.4~ \pm ~\,0.8$ & 6.5  $+1.0 \choose -1.1$ \nl
1984L & Ib & $21.7~ \pm ~\,3.3$   & 19.1 $+{\rm ind.} \choose -12.1$ \nl
1990B & Ic & $19.4~ \pm ~\,2.9$   & 20.7 $+3.2 \choose -3.3$ \nl
1994I & Ic & $~\,8.9~ \pm ~\,1.3$ &  6.1 $+1.0 \choose -0.9$ \nl
\sidehead{Type II}
1970G & IIL & $~\,7.4~ \pm ~\,0.6$ & 7.4 $+0.7 \choose -1.4$ \nl
1978K & II & $~\,4.5~ \pm ~\,0.7$  & 3.9 $+0.8 \choose -0.8$ \nl
1979C & IIL & $17.1~ \pm ~\,1.8$  & 20.1 $+2.1 \choose -2.2$ \nl
1980K & IIL & $~\,6.3~ \pm ~\,1.0$ & 12.7 $+2.0 \choose -1.9$ \nl
1981K & II & $~\,5.9~ \pm ~\,1.1$ & 3.3 $+{\rm ind.} \choose -1.1$ \nl
1985L & IIL & $~\,21.6~ \pm ~\,3.2$ & 42.5 $+7.2 \choose -9.8$ \nl
1986E & IIL & $~\,19.4~ \pm ~\,2.9$ & 49.0 $+8.3 \choose -8.3$ \nl
1986J & IIn & $~\,11.1~ \pm ~\,1.7$ & 7.2 $+1.1 \choose -1.1$ \nl
1987A & IIpec  & $~\,0.051~ \pm ~\,0.002$ & 0.071 $+0.1 \choose -0.1$ \nl
\tablebreak
1988Z & IIn & $102.6~ \pm 15.4$ & 70.7 $+10.7 \choose -10.6$ \nl
1993J & IIb & $~\,3.6~ \pm ~\,0.3$ & 2.4 $+0.2 \choose -0.2$ \nl
1996cb & IIb & $~\,9.1~ \pm ~\,1.4$ & 3.8 $+0.7 \choose -0.6$ \nl
\enddata 
\tablenotetext{a}{From Table 1.}  
\tablenotetext{b}{Distance
which the RSN would have if: 1) for Type Ib/c RSNe, the peak observed
6 cm flux density originates from a ``standard candle'' radio source
with peak luminosity given, respectively, by Eqs. 15 and 16; or 2)
for Type II RSNe, the peak observed 6 cm flux density originates from
a source with peak luminosity given by Eq. 17, with time in days.}
\end{deluxetable}

\clearpage

\begin{figure} 
\plotfiddle{distfig1.eps}{300pt}{0}{75}{75}{-225}{-110}
\caption{Schematic representation of the competing effects of rapidly
declining absorption ({\it long-dashed curve}) and more slowly
declining emission ({\it short-dashed curve}), which yield the
characteristic rapid turn-on, slower turn-off ``light curve'' ({\it
solid curve}) observed for RSNe at any given frequency.}
\label{fig1}
\end{figure} 

\clearpage

\begin{figure} 
\plotfiddle{distfig2.eps}{300pt}{0}{75}{75}{-225}{-110}
\caption{Schematic representation of the radio light curves at a given
frequency for SNe with differing mass-loss rate/presupernova stellar
wind velocity (${\dot M / w}$) ratios.  Note that the longer the
time delay required to reach peak luminosity the higher is that
luminosity peak.}
\label{fig2}
\end{figure} 

\clearpage

\begin{figure}
\plotfiddle{distfig3.eps}{300pt}{-90}{65}{65}{-250}{+400}
\caption{Peak 6 cm luminosity, $L_{\rm 6\ cm\ peak}$ of RSNe {\it
vs.\/} time, in days, from explosion to peak 6 cm flux density
($t_{\rm 6\ cm\ peak} - t_0$).  Type II SNe are plotted as {\it filled
triangles}.  The {\it dashed line\/} is the unweighted, best fit to
the 12 available Type II RSNe.  The Type II SNe show a large range in
times to 6 cm peak ($t_{\rm 6\ cm\ peak} - t_0$) and in peak 6 cm
luminosity $L_{\rm 6\ cm\ peak}$, but appear to obey the relation
given by Eq.~17.  Error bars are based on best estimates (see
$\S$3.2).  Where no error or only a stub of a line is shown, the error
in that direction is indeterminate.}
\label{fig3}
\end{figure} 


\begin{thebibliography}{}

\bibitem[Allen et al.~1976]{all76} Allen, R.~J., Goss, W.~M., Ekers,
R.~D., \& de Bruyn, A.~G. 1976, A\&A, 48, 253
\bibitem[Chevalier 1982a]{che82a} Chevalier, R.~A. 1982a, ApJ, 259,
    302
\bibitem[Chevalier 1982b]{che82b} Chevalier, R.~A. 1982b, ApJ, 259,
    L85
\bibitem[Chevalier 1984]{che84} Chevalier, R.~A. 1984, ApJ, 285, L63
\bibitem[Cowan, Goss, \& Sramek 1991]{cow91} Cowan, J.~J., Goss,
W.~M., \& Sramek, R.~A. 1991, ApJ, 379, L49
\bibitem[Freedman et al.~1994a]{fre94a} Freedman, W.~L. et al.~1994a,
Nature, 371, 757
\bibitem[Freedman et al.~1994b]{fre94b} Freedman, W.~L. et al.~1994b,
ApJ, 427, 628
\bibitem[Greenhill et al.~1995]{gre95} Greenhill, L.~J., Jiang, D.~R.,
Moran, J.~M., \& Reid, M.~J. 1995, ApJ, 440, 619
\bibitem[Hamuy et al.~1996]{ham96} Hamuy, M., Phillips, M.~M.,  Suntzeff, N.~B., Schommer, R.~A., Maza, J., \& Aviles, R.~1996, AJ, 112, 2398
\bibitem[Kelson et al. ~1996]{Kel96} Kelson, D.~D. et al. 1996, ApJ, 463, 26
\bibitem[Marscher \& Brown 1978]{mar78} Marscher, A.~P., \& Brown,
R.~L. 1978, ApJ, 220, 474
\bibitem[Miyoshi et al.~1995]{miy95} Miyoshi, M., Moran, J.,
Herrnstein, J., Greenhill, L., Nakai, N., Diamond, P., \& Inoue,
M. 1995, Nature, 373, 127
\bibitem[Montes et al. ~1997]{mont197} Montes, M.~J., Van Dyk, S.~D.,
Weiler, K.~W. Sramek, R.~A., \& Panagia, N. 1997, ApJ, 482, L61
\bibitem[Montes, Weiler, \& Panagia 1997]{mont297} Montes, M.~J.,
Weiler, K.~W. \& Panagia, N. 1997, ApJ, 488, 792
\bibitem[Panagia 1982]{pan82} Panagia, N. 1982, Ultraviolet Properties
of Supernovae. {\it In ESA 3rd European IUE Conf.}, pp. 31-36, ESA
\bibitem[Panagia, Sramek, \& Weiler 1986]{pan86} Panagia, N., Sramek,
R.~A., \& Weiler, K.~W. 1986, ApJ, 300, L55
\bibitem[Panagia, et al.~1991]{pan91} Panagia, N., Gilmozzi, R.,
 Macchetto, F., Adorf, H. -M.,  Kirshner, R.~P.~1991, ApJ, 380, L23
\bibitem[Press et al.~1992]{pre92} Press, W.~H., Teukolsky, S.~A., Vetterling, W.~T., Flannery, B.~P. 1992, Numerical Recipes in FORTRAN: The Art of Scientific Computing, Cambridge Univ.~Press
\bibitem[Riees, Press, \& Kirshner~1996]{rie96} Riess, A.~G., Press, W.~H., \& Kirshner, R.~P.~1996, ApJ, 473, 88
\bibitem[Rupen et al.~1998]{rup98} Rupen, M.~P. et al.~1998, in
preparation
\bibitem[Ryder et al.~1993]{ryd93} Ryder, S., Staveley-Smith, L.,
Dopita, M., Petre, R., Colbert, E., Malin, D., \& Schlegel, E. 1993,
ApJ, 417, 167
\bibitem[Saha et al.~1997]{sah97} Saha, A., Sandage, A.,
 Labhardt, L., Tammann, G.~A.,  Macchetto, F.~D., \& Panagia, N.~1997, ApJ, 486, 1
\bibitem[Schmidt, Kirshner, \& Eastman 1992]{sch92} Schmidt, B.~P.,
Kirshner, R.~P., \& Eastman, R.~G. 1992, ApJ, 395, 366
\bibitem[Schmidt et al.~1994]{sch94} Schmidt, B.~P., Kirshner, R.~P.,
Eastman, R.~G., Phillips, M.~M., Suntzeff, N.~B., Hamuy, M., Maza, J.,
\& Aviles, R.  1994, ApJ, 432, 42
\bibitem[Sramek, Panagia, \& Weiler 1984]{sra84} Sramek, R.~A.,
Panagia, N., \& Weiler, K.~W. 1984, ApJ, 285, L59
\bibitem[Stathakis \& Sadler 1991]{sta91} Stathakis, R.~A., \& Sadler,
E.~M.  1991, MNRAS, 250, 786
\bibitem[Tully 1988]{tul88} Tully, R.~B. 1988, Nearby Galaxies
Catalogue, Cambridge Univ.~Press
\bibitem[Turtle 1987]{turt87} Turtle, A.~J., Campbell-Wilson, D.,
Bunton, J.~D., Jauncey, D.~L., Kesteven, M.~J., Manchester, R.~N.,
Norris, R.~P., Storey, M.~C., \& Reynolds, J.~E. 1987, Nature, 327, 38
\bibitem[van der Hulst et al.~1983]{van83} van der Hulst, J.~M.,
Hummel, F., Davies, R.~D., Pedlar, A., \& van Albada, G.~D. 1983,
Nature, 306, 566
\bibitem[Van Dyk et al.~1992]{van92} Van Dyk, S.~D., Weiler, K.~W.,
Sramek, R.~A., \& Panagia, N. 1992, ApJ, 396, 195
\bibitem[Van Dyk et al.~1993a]{van93a} Van Dyk, S.~D., Sramek, R.~A.,
Weiler, K.~W., \& Panagia, N. 1993a, ApJ, 409, 162
\bibitem[Van Dyk et al.~1993b]{van93b} Van Dyk, S.~D., Sramek, R.~A.,
Weiler, K.~W., \& Panagia, N. 1993b, ApJ, 419, L69
\bibitem[Van Dyk et al.~1994]{van94} Van Dyk, S.~D., Weiler, K.~W.,
Sramek, R.~A., Rupen, M.~P., \& Panagia, N. 1994, ApJ, 432, L115
\bibitem[Van Dyk et al.~1996]{van96} Van Dyk, S.~D., Sramek, R.~A.,
Montes, M.~J., Weiler, K.~W., \& Panagia, N. 1996, IAUC 6528
\bibitem[Van Dyk et al.~1998]{van98} Van Dyk, S.~D., Montes, M.~J.,
Weiler, K.~W., Sramek, R.~A., \& Panagia, N. 1998, \apj, in press
\bibitem[Weiler et al.~1981]{wei81} Weiler, K.~W., van der Hulst,
J.~M., Sramek, R.~A., \& Panagia, N. 1981, ApJ, 243, L151
\bibitem[Weiler et al.~1986]{wei86} Weiler, K.~W., Sramek, R.~A.,
Panagia, N., van der Hulst, J.~M., \& Salvati, M. 1986, ApJ, 301, 790
\bibitem[Weiler, Panagia, \& Sramek 1990]{wei90} Weiler, K.~W.,
Panagia, N., \& Sramek, R.~A. 1990, ApJ, 364, 611
\bibitem[Weiler et al.~1991]{wei91} Weiler, K.~W., Van Dyk, S.~D.,
Panagia, N., Sramek, R.~A., \& Discenna, J.~L. 1991, ApJ, 380, 161
\bibitem[Weiler et al.~1992]{wei92} Weiler, K.~W., Van Dyk, S.~D.,
Panagia, N., \& Sramek, R.~A. 1992, ApJ, 398, 248

\end{thebibliography}
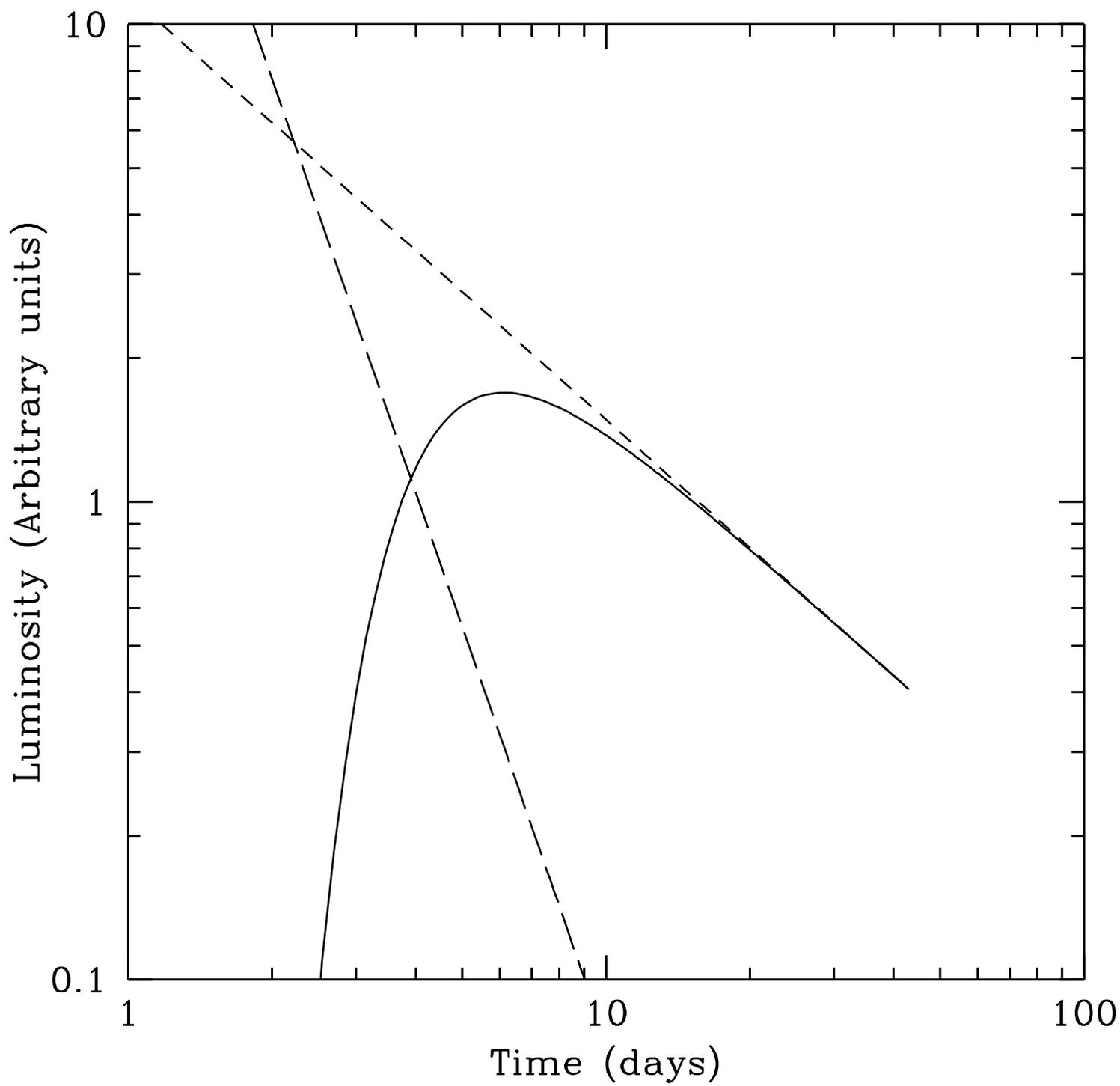
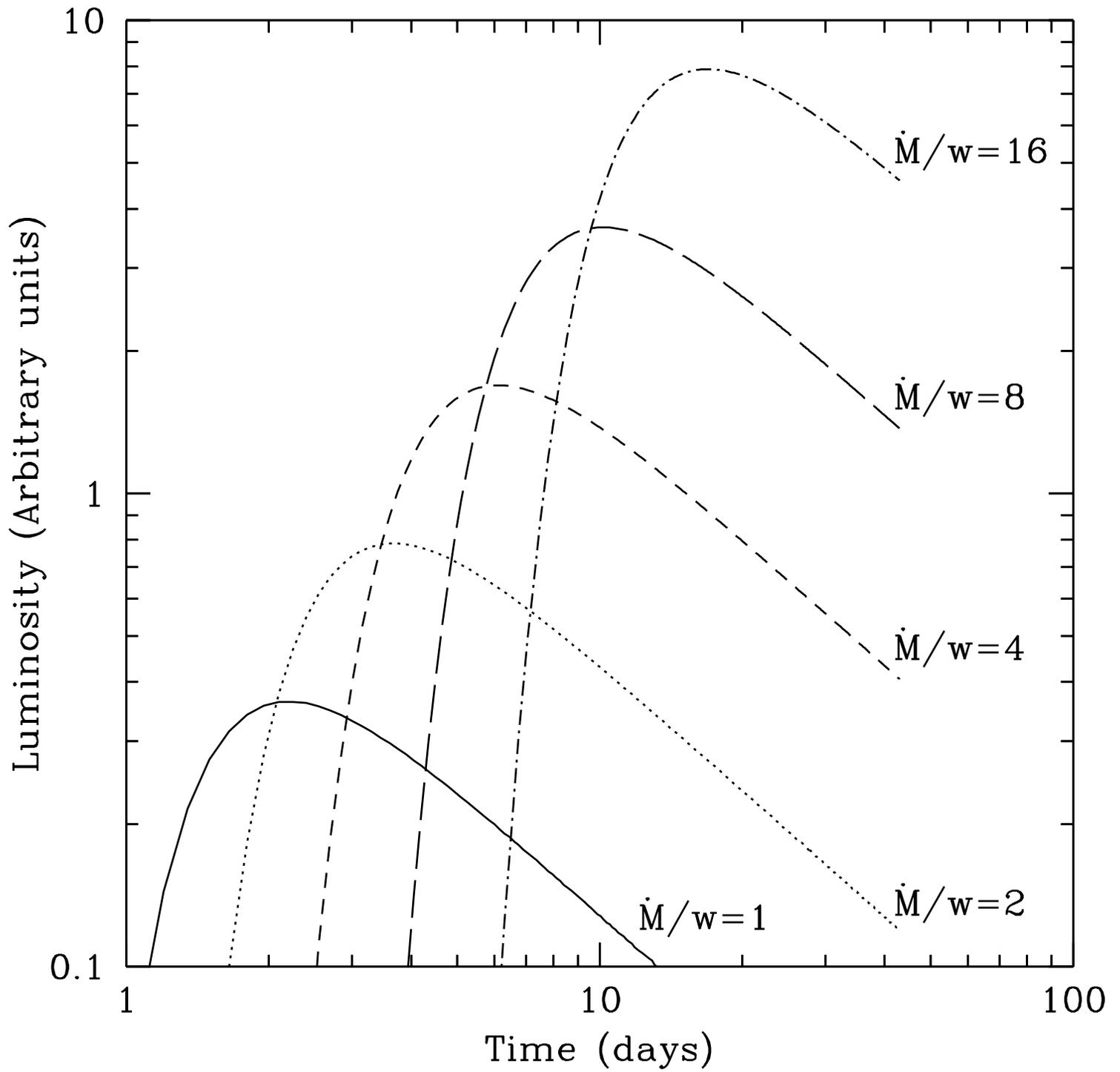
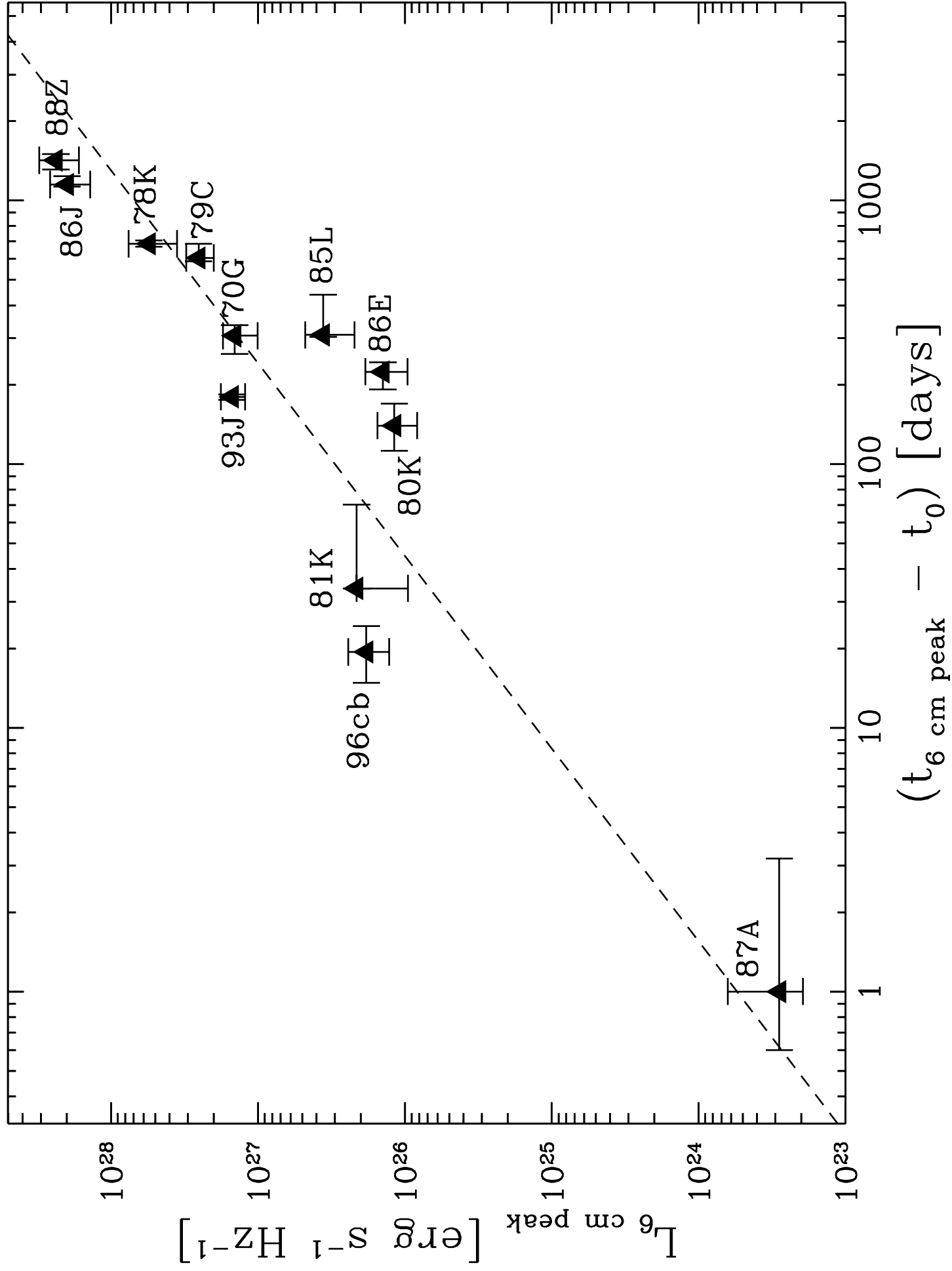
\end{document}